\documentclass[preprint2]{aastex}

\lefthead{Escala \& Larson}
\righthead{Stability of Galactic  Gaseous Disks and the Formation of Massive Clusters.}

\def\msun{M_{\odot}}

\begin{document}

\title{Stability of Galactic Gas Disks and the Formation of Massive Clusters.}

\author{Andr\'es Escala}
\affil{Kavli Institute for Particle Astrophysics and Cosmology}
\affil{Stanford University Physics Department / SLAC, 2575 Sand Hill Rd. MS 29, Menlo Park, CA 94025, USA.}
\author{Richard B. Larson}
\affil{Department of Astronomy, Yale University, New Haven, CT 06520-8101, USA.}

\begin{abstract}

We study gravitational instabilities in disks, with special attention to the most massive clumps that form because they are expected to be the progenitors of globular-type clusters.  The maximum unstable mass is set by rotation and depends only on the surface density and orbital frequency of the disk. We propose that the formation of massive clusters is related to this largest scale in galaxies not stabilized by rotation. Using data from the literature, we predict that globular-like clusters can form in nuclear starburst disks and protogalactic disks but not in typical spiral galaxies, in agreement with observations. 
\end{abstract}


\section{Introduction}

The study of instabilities in disks has a long history, following the seminal work of Toomre (1964), with a considerable literature on many aspects of it.  However, relatively little attention has been given to the most massive agglomerations that can form by the fragmentation of galactic gas disks.  These most massive agglomerations are of interest because they may be the precursors of the most massive star clusters known, the globular clusters.

Globular clusters were until recently viewed as exclusively old objects, and cluster formation models were therefore based on ideas about early stages of galaxy formation.  This view changed with the realization that elliptical galaxies often contain two populations of globular clusters that appear to have different origins, suggested to be a `primordial' population and a `merger' population (Ashman \& Zepf 1992).  The existence of a merger population was confirmed by the Hubble Space Telescope with the discovery of massive young clusters in merging galaxies (Ashman \& Zepf 1998; Schweizer 1998).  A modern theory of globular cluster formation should therefore address why globular cluster formation is common in some environments such as merging and high-redshift galaxies but not in others such as the present Milky Way.

Another question to be addressed is why some regions of galaxies are more favorable for the formation of massive clusters than others.  For example, the inner part of the Milky Way galaxy contains young clusters with masses up to several times $10^4\,\msun$, including some quite close to the central black hole (Krabbe et al. 1995; Figer 2008), that are an order of magnitude more massive than the typical open clusters found elsewhere in the galaxy.

The Giant Molecular Clouds (GMCs) in which open clusters currently form in the outer Milky Way have masses of order $10^6\,\msun$, but they produce clusters with masses of only several times $10^3\,\msun$.  The formation of globular clusters with masses of $10^5 - 10^6\,\msun$ therefore requires either much more massive super-GMCs (Harris \& Pudritz 1994) or a much higher star formation efficiency.  In reality, a combination of these effects is probably involved. 
 
Several scenarios have been proposed for the formation of massive super-GMCs, involving for example a hot primordial plasma (Fall \& Rees 1985), collisions between normal GMCs (Harris \& Pudritz 1994), or confinement by high pressures in mergers (Ashman \& Zepf 2001).  Relatively little attention has been given to the possible role of disks in the formation of globular clusters, with the exception of Larson (1988, 1996), but this possibility now seems worth further investigation given the evidence for massive cluster formation in rotationally supported disks in merging galaxies (Ashman \& Zepf 1998).

In this $Letter$ we study the gravitational instability of galactic gas disks, with special attention to the most massive clumps that can form in them.  We also consider which galaxies or environments will produce the largest unstable clumps and are therefore most favorable for massive cluster formation.  We start by reviewing the theory of instabilities in disks in \S 2, and continue with its application to  the formation of globular-type clusters in \S 3.  We discuss the most promising environments for globular-cluster formation in \S 3.1, and in \S 4 we summarize the main implications of our work.

\section{GRAVITATIONAL INSTABILITY IN DISKS}

For illustration we consider the simplest possible case: a uniformly rotating thin sheet or disk.  Linear stability analysis of such a system yields the  dispersion relation for small perturbations in surface density (Binney \& Tremaine 1987)  
\begin{equation}
\omega^2 = 4 \Omega^2 - 2\pi G \Sigma |k| + k^2 C_{\rm s}^2 \,,
\label{dispersion}
\end{equation} 
where $C_{\rm  s} = \sqrt\frac{dP}{d\Sigma}$ is the sound speed, $\Sigma$ is the surface density, and $\Omega$ is the angular rotation speed.  To see the implications of equation \ref{dispersion}, we first consider the limit of a non-rotating sheet with $\Omega = 0$. In this case the sheet is unstable ($\omega^2 < 0$) for wavelengths $\lambda = 2\pi / k$ larger than  the Jeans length, $\lambda_{\rm Jeans} = C_{\rm s}^2 / G\Sigma $.  Thus pressure can only stabilize the sheet at small scales.  Another limiting case is a pressureless sheet with $C_{\rm  s} = 0 $.  Equation \ref{dispersion} then shows that only perturbations with $\lambda < \lambda_{\rm rot} = \pi^2 G \Sigma / \Omega^2$ are unstable, so that rotation can only stabilize the sheet at large scales. The existence of a maximum length scale not stabilized by rotation was first derived and discussed by Toomre (1964).

Clearly neither pressure nor rotation can by itself stabilize the sheet, and there is a range of unstable length scales limited on small scales by thermal pressure (at the Jeans length $\lambda_{\rm Jeans}$) and on large scales by rotation (at the critical length set by rotation, $\lambda_{\rm rot}$).  All intermediate length scales are unstable, and the most rapidly growing mode has a wavelength $2 \, \lambda_{\rm Jeans} $.  Only a combination of pressure and rotation can stabilize the sheet, and this happens when the range of unstable wavelengths shrinks to zero; according to equation \ref{dispersion} this occurs when $\lambda_{\rm Jeans} \geq \lambda_{\rm rot}/4$, which is equivalent to $C_{\rm s}\Omega/ G \Sigma_{0} \ge \pi/2$, the classical Toomre (1964) criterion for the stability of a uniformly rotating disk.

Such stability analyses apply only to fluid disks that are well described by a simple equation of state (EOS), such as an isothermal or a polytropic EOS.  Unfortunately, the real interstellar medium in galaxies is highly complex and not well described by a simple equation of state.  The complex dynamics and thermodynamics of the real interstellar medium produce large fluctuations and sharp transitions in its properties, as happens for example in molecular clouds, that change the Jeans mass by orders of magnitude and are not well described by a simple EOS.  A medium with complex structure and dynamics may therefore have no well-defined Jeans length, and there may be no real lower limit on the sizes of the self-gravitating structures that can form until the much smaller thermal Jeans scale is reached in molecular cloud cores.  This thermal Jeans scale may be the next smaller scale that has any clear physical basis and where a relatively simple EOS may again apply; stability analyses give length and mass scales in molecular cloud cores of around 0.1 pc and 1 $\msun$. 

The maximum length scale set by rotation $\lambda_{\rm rot}$ is however still a meaningful quantity even when the ISM is not well described by a simple EOS, because $\lambda_{\rm rot}$ depends only on the surface density and angular frequency of the disk and not on its small-scale physics.  If a disk has a well-defined average surface density, it should have a well-defined maximum length scale for instabilities.  This rotational maximum scale may then be the most relevant length scale for the large-scale dynamics of a disk with complex small-scale physics.  In the Milky Way, $\lambda_{\rm rot}$ is typically of the order of 1 kpc.  At intermediate scales between 0.1 pc and 1 kpc there may not be any preferred scale, and it has been suggested by Larson (1979, 1981) and others (Ballesteros-Paredes et al. 2007 and references therein) that the structure and dynamics of the ISM on these intermediate scales may be roughly self-similar and described by power laws, as in a turbulent cascade.

Since the maximum unstable length scale set by rotation is the only clearly meaningful length scale for a gas disk with complex physics, we focus on the  characteristic mass associated with this scale.  The resulting maximum unstable mass, calculated as $M^{\rm max}_{\rm cl} = \Sigma_{\rm gas} \, (\lambda_{\rm rot}/2)^2$, is
\begin{equation}
M^{\rm max}_{\rm cl} = \frac{\pi^4G^2\Sigma_{\rm gas}^3}{4\Omega^4} \, .
\label{maxcl}
\end{equation}
For typical parameters in the Milky Way disk, equation \ref{maxcl} gives a maximum unstable mass of the order of a few times $10^6\,\msun$, similar to the scale of giant cloud complexes in our Galaxy.  Equation \ref{maxcl} gives a mass scale similar to that derived previously by many authors (Balbus 1988; Elmegreen 1994, 2002; Kim \& Ostriker 2001) on the assumption that it is the Jeans mass $M_{\rm Jeans} = \Sigma_{\rm gas} (\lambda_{\rm Jeans}/2)^2$ that determines the masses of giant cloud complexes, where the Jeans mass is calculated assuming that on large enough scales the ISM can be modeled as an isothermal fluid in which random cloud motions take the place of thermal motions.  This mass scale is again about $10^6 - 10^7\,\msun$, and it is similar to the rotational maximum mass because the analysis assumes that disks are marginally stable ($Q \sim 1$), implying that the Jeans scale and the rotational scale are similar.

A significant radial variation is implied by  \ref{maxcl}, since both $\rm \Sigma_{\rm gas}$ and $\rm \Omega$ are  functions of radius in galactic disks. For   disks with  flat rotation curves that have  density profiles steeper than $\rm \Sigma_{\rm gas} \propto R^{-4/3}$, the maximum clump mass decreases with radius. For example in the Milky Way, the maximum  mass  decreases with radius from about $\rm 10^{6} \, \msun$ at the solar radius  to $\rm 3\times 10^{4} \, \msun$ at 20 kpc from the galactic center.  

In the more general case of non-uniform rotation, equations \ref{dispersion} and \ref{maxcl} are easily generalized by replacing $2\Omega$ by  the epicyclic frequency $\kappa$, which is given by $\kappa^2 (R) = R \frac{d\Omega^2}{dR} \, + \, 4 \Omega^2$.  Since  the epicyclic frequency  $\kappa$ varies only between $\Omega$ and $2\Omega$ for centrally concentrated disks, we will continue to use $\Omega$ because it has a more simple and intuitive meaning.  For a review of results of more general gravitational stability analyses see Larson (1985). 

The effect of rotation on the properties of the massive clumps that form in disks is illustrated by the simulations of nuclear gas disks by Escala (2007).  These simulations study the evolution of a gas disk with a mass of $3 \times 10^8\,\msun$ and a diameter of 600 pc, varying only the constant rotation velocity of the disk and studying its effect on the disk's evolution.  This was done by varying a fixed external potential in such a way that the disk is always rotationally supported, but with a range of rotation velocities.  Here we compare the development of the instabilities in two disks with rotational velocities $v_{\rm rot}$ of 110 and 321 $\rm km\,s^{-1}$.  Further details of the simulations are given by Escala (2007).

\begin{figure}
\plotone{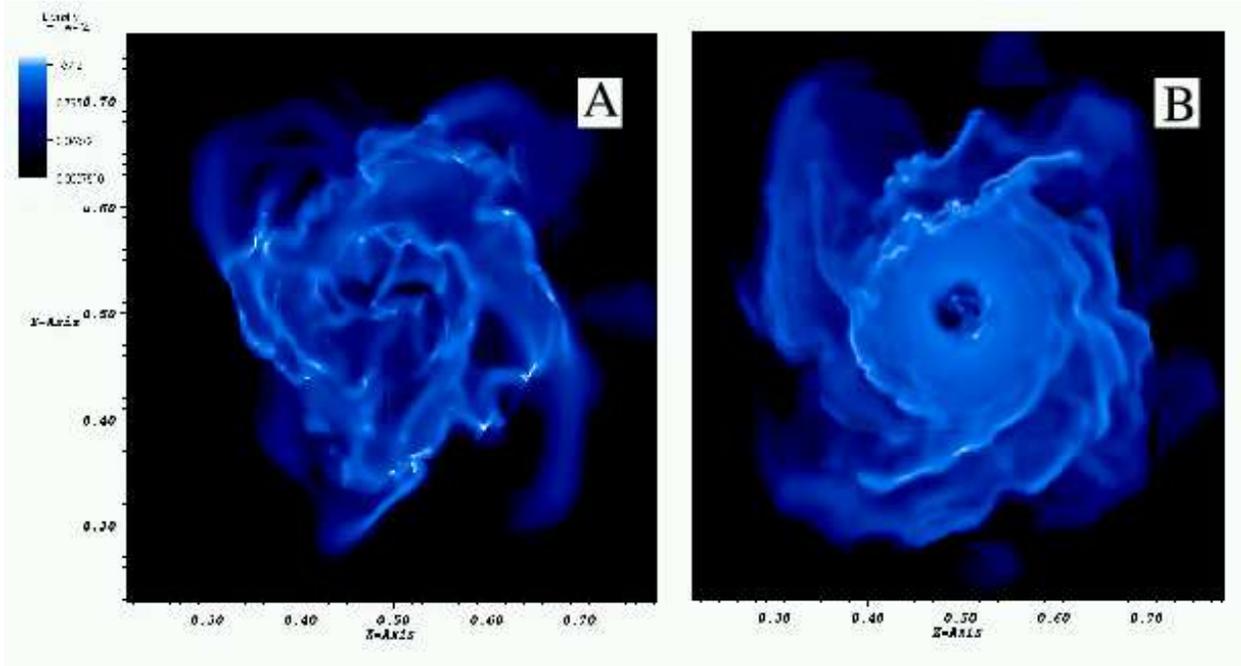} 
\caption{a) Density distribution in the plane of the gas disk, in units of 1.19 $\rm \times 10^{-21}$ g $\rm cm^{-3}$ and coded on a logarithmic scale, at time t = 0.4 $\rm t_{orb}$  for  the disk with rotational velocity   of  110  $\rm km \, s^{-1}$. The plotted region in the x-y plane is 0.5\,kpc $\times$ 0.5\, kpc. The figure shows a complicated multiphase   structure,  characterized by high density clumps that are embedded in a less dense medium. b) Same as in a) for the disk with rotational velocity   of 321  $\rm km \, s^{-1}$.  The figure also shows a complicated multiphase   structure as in a), but now characterized by  filaments  that are embedded in a less dense medium}
\label{disk}
\end{figure}

Figure $\ref{disk}$ illustrates a representative stage in the evolution of the two disks and shows the density distribution in the disk midplane after 0.4 orbital periods.  The disk with a rotational velocity $v_{\rm rot}$ of 110  $\rm km\,s^{-1}$ is shown in Figure $\ref{disk}$a and the disk with $v_{\rm rot}$ = 321 $\rm km\,s^{-1}$ in Figure $\ref{disk}$b.  The plotted region in the x-y plane is 0.5\,kpc on a side.  Both disks develop a complex clumpy structure, and the figures illustrate how rotation affects the growing instabilities.  In the run with lower rotation (Fig.\ $\ref{disk}$a) the clumps are denser and more numerous, while in the more rapidly rotating disk (Fig.\ $\ref{disk}$b) the clumps are less marked and filaments are more prominent, reflecting the disruption of some clumps by rotation.  This is an expected result since Figure $\ref{disk}$b has a smaller $M^{\rm max}_{\rm cl}$, so that rotation can here disrupt clumps that are able to survive in Figure $\ref{disk}$a; this explains why the clumps are denser and more numerous in the run with less rotation.

The gas clumps in Figure $\ref{disk}$a have typical masses of several times $10^6\,\msun$, in rough agreement with the maximum clump mass of $1.5 \times 10^7\,\msun$ predicted by equation \ref{maxcl}.  For the disk with higher rotation (Fig.\ $\ref{disk}$b), the clumps have smaller typical masses of several times $10^5\,\msun$, also in agreement with  the value of $4 \times 10^5\,\msun$ predicted by equation \ref{maxcl}.  These values are valid for the early evolution of the disk before much of its gas has been consumed by star formation.  At later times, as $\Sigma_{\rm gas}$ decreases, less massive clumps are formed.

\section{Applications: The largest proto-globular clusters}

The condition for rotational support of a homogeneous uniformly rotating gas sheet against the gravity produced by the total mass enclosed within any radius $R$, $M_{\rm tot} = M_{\rm star} +  M_{\rm DM} + M_{\rm gas} $, can be written
\begin{equation}
\Omega^2 = \frac{\pi G \Sigma_{gas}}{\eta \, R} \, ,  
\end{equation}
where $\eta = M_{\rm gas}/M_{\rm tot}$ is the ratio of the gas mass to the total enclosed mass.  For a uniformly rotating, rotationally supported gas sheet, the maximum clump mass predicted by equation \ref{maxcl} can be written
\begin{equation}
M^{\rm max}_{\rm cl} = \frac{\pi^2 \eta^2 R^2 \Sigma_{\rm gas}}{4} = 3\,\, 10^7\,\msun \,\, \frac{ M_{\rm gas}}{10^9\,\msun} \left(\frac{\eta}{0.2}\right)^2.
\label{eq4}
\end{equation}
Equation \ref{eq4} shows that the formation of massive clusters requires not only a large gas content, as in the Milky  Way, but also a large gas fraction (large $\rm \eta$); the formation of massive clusters is thus strongly favored in regions where the gas constitutes a large fraction of the mass.

In the local universe this condition is satisfied in ULIRGs, which are typically characterized by an ongoing burst of star formation.  Recent observations have shown that massive clusters are in fact currently forming in these systems (Ashman \& Zepf 1998).  In these starburst systems, not all of the star formation is confined to a single well-defined disk, especially during the early stages of a galaxy merger before it settles down into an ordered flow, as in the Antennae system.  However, the maximum mass derived above is due to a purely local effect and not dependent on large-scale ordered motion. The mass scale is set  by rotation via the local balance between centrifugal and self-gravitational forces, and it is not strongly sensitive to the geometry; equation \ref{maxcl} should therefore remain approximately valid within geometrical factors.  For a typical nuclear starburst disk with a mass of a few times $10^9\,\msun$ and a gas fraction $\eta$ of 1/5 (Downes \& Solomon 1998), we predict the formation of clumps with masses up to $10^8\,\msun$.  The formation efficiency of massive clusters is very uncertain, but even with an efficiency of only 1\% we predict the formation of clusters with masses up to $10^6\,\msun$, characteristic of the most massive globular clusters.

In normal disk galaxies like the Milky Way, the total gas mass is similar but the gas fraction is only about 1/30 of the total mass.  We then predict clumps with masses of order a few times $10^6\,\msun$, as mentioned in \S2, that are not massive enough to form globular clusters.  However, conditions for massive cluster formation become much more favorable in the protogalactic gas disks that are expected to exist during the early stages of disk galaxy formation, where the gas is expected to constitute 1/5 or more of the total mass ($\eta \sim 0.2$).  Such protogalactic disks, like for example the clump cluster and chain galaxies observed at high redshift (Elmegreen \& Elmegreen 2005), would then also be very favorable sites for massive cluster formation.

Thus we see from equation \ref{eq4} that there are two favorable environments for the formation of massive clusters: (1) nuclear gas disks in starburst systems, and (2) protogalactic disks.  This expectation agrees with the scenario for globular cluster formation proposed by Ashman \& Zepf (1992), whereby globular clusters form either in the early stages of galaxy formation or in galaxy mergers and interactions.  This scenario has been successfully tested against several observed properties of globular cluster systems, such as the typically bimodal age and metallicity distributions of globular clusters in elliptical galaxies and the presence of young globular clusters in merger remnants. Moreover it has been recently found that  the oldest dwarf galaxies have the highest globular cluster frequencies (Peng et al. 2008), which is expected since  systems that formed earlier would probably  have had higher gas fractions and therefore they were more favorable sites for  cluster formation.

\subsection{Environments for Massive Cluster Formation}

\begin{figure}
\plotone{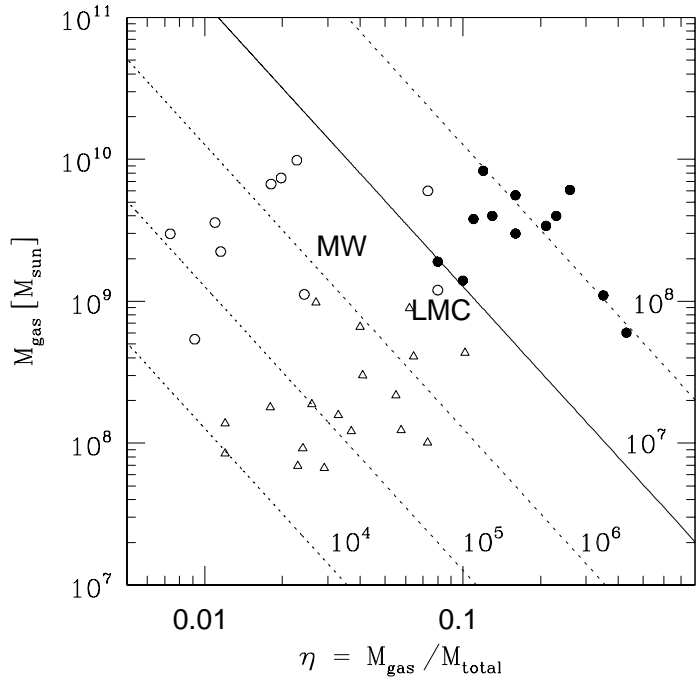} 
\caption{Total gas mass in a galaxy plotted against   the gas fraction $M_{\rm gas}/M_{\rm tot}$. The black dots are nuclear starburst  disks, open circles are spirals and triangles are dwarf irregulars. The parallel dotted lines indicate predicted clump masses from equation \ref{eq4} of $10^4\,\msun$, $10^5\,\msun$, $10^6\,\msun$ and $10^8\,\msun$. The solid line is for $10^7\,\msun$. MW  denotes the position of the Milky  Way in the plot, and LMC the position of the  Large  Magellanic Cloud.}

\label{f1}
\end{figure}

The formulation presented in this {\it Letter} also allows us to predict which present environments are most favorable for forming massive clusters, given the total gas mass and the gas fraction in galaxies, quantities that can easily be extracted from observations of many systems.  Figure \ref{f1} shows the total gas mass of a sample of galaxies plotted against the gas fraction $\eta = M_{\rm gas}/M_{\rm tot}$, where the black dots are nuclear starburst disks, the open circles are spirals, and the triangles are dwarf irregular galaxies.  The data for spiral galaxies are from Kent (1987), Giraud (1998), and Helfer et al. (2003).  For  dwarf irregular galaxies the data are from C\^ ot\' e et al. (2000), Van den Bosch et al. (2001), and for starburst galaxies they are from Downes and Solomon (1998). The parallel dotted lines show the maximum clump mass predicted by equation \ref{eq4}, with values varying from $10^4\,\msun$ to $10^8\,\msun$; the solid line is for a maximum clump mass of $10^7\,\msun$.

We expect globular clusters to form in clumps with masses above $10^7\,\msun$, so that galaxies above the solid line in Figure \ref{f1} should be good candidates for massive cluster formation.  If massive clusters form with an efficiency of 1\% in GMCs (Lada \& Lada 2003), this corresponds to cluster masses of $10^5\,\msun$ or more, comparable to globular cluster masses.  Figure \ref{f1} shows that with one exception, only galaxies classified as nuclear starbursts lie on or above the $10^7\,\msun$ line, in agreement with the fact that only starburst galaxies show ongoing massive cluster formation.  The one exception is the  spiral galaxy NGC 2403, which has many giant HII regions and is forming massive stellar clusters like starbursts (Drissen et al 1999), in agreement with our prediction.

Figure \ref{f1} also shows that dwarf irregular galaxies should not form massive clusters even though most of their baryonic mass is in gas.  This is because the more important quantity is the gas fraction, which is only between 1\% and 10\% in these systems because they are dark-matter dominated.  In spiral galaxies, although the total amount of gas is comparable to that in starburst galaxies, the gas is more widely distributed and the gas fraction in the star-forming region is therefore smaller, so we do not expect massive cluster formation in most spirals, in agreement with observations. 

The position of the Milky Way in Figure \ref{f1} is denoted by MW and the position of the Large Magellanic Cloud is denoted by LMC.  Both the Milky Way and the LMC are predicted to have maximum clump masses of several times $10^6\,\msun$, which for a cluster formation efficiency of 1\% is consistent with the fact that they are both currently forming clusters with masses up to several times $10^4\,\msun$ (see Figer 2008 for the Milky Way and Larson 1988 for LMC).  However, given the uncertainty in the cluster formation efficiency, what is more significant is the prediction that the Milky Way and the LMC should form clusters two orders of magnitude less massive than the most massive clusters that form in starbursts, as is observed.

\section{DISCUSSION}

In this paper we have studied gravitational instabilities in disks, with special attention to the most massive clumps that form because they are expected to be the progenitors of globular-type clusters.  The maximum unstable mass is set by rotation and depends only on the surface density and orbital frequency of the disk, unlike other mass scales such as the Jeans mass that depend on the complex gas physics. This mass scale is due to a purely local instability and not dependent on large-scale ordered motion. The maximum unstable mass is therefore a well-defined quantity even if the interstellar medium is not well described by a simple equation of state.

The maximum clump mass can be expressed in terms of the total gas mass and the gas fraction in a galaxy, and this formulation makes it clear that environments with a high gas fraction are the most promising places to form massive clusters.  Using data from the literature, we predict that massive globular-like clusters can form in nuclear starburst disks and protogalactic disks but not in typical spiral galaxies, in agreement with observations.

The scenario proposed here relates massive clusters to the largest scale in galaxies not stabilized by rotation, which is the only scale intermediate between stars and galaxies that has a clear physical basis.  There is no well-defined `Jeans mass' on these intermediate scales, and the next smaller scale that has any clear physical basis is the thermal Jeans scale in molecular clouds, which is related to the masses of individual stars.  The next larger physical scale is that of the galaxy itself, so we predict three physically well-motivated scales corresponding to stars, massive stellar  clusters, and galaxies.   

Another application of studies of the stability of nuclear gas disks, not discussed here, is to their likely role in feeding supermassive central black holes in galaxies (Escala 2007).  This requires the outward transfer of angular momentum, and if gravity is the most important force involved, the same mass concentrations that form massive clusters may also play an important role in the outward transfer of angular momentum and the growth of a central black hole.  We plan to investigate this problem further in ongoing work.

\onecolumn

\end{document}